\documentstyle[12pt,preprint]{aastex}

\newcommand{\et}{et al.}

\newcommand{\ha}{H$\alpha$}
\newcommand{\solar}{\ifmmode_{\sun}\;\else$_{\sun}\;$\fi}

\newcommand{\HII}{H$\,${\sc ii}}
\newcommand{\HI}{H$\,${\sc i}}

\newcommand{\x}{\enspace}

\newcommand{\sfrunit}{M\solar\ yr$^{-1}$ kpc$^{-2}$}

\begin{document}

\title{Power Spectra in V-band and \ha\ of 9 Irregular Galaxies}

\author{Kyle W. Willett\footnote{Current affiliation: Dept. of Physics
and Astronomy, Carleton College, 300 N. College St., Northfield, Minnesota
55057}}
\affil{Lowell Observatory, 1400 West Mars Hill Road, Flagstaff, Arizona
86001 USA}
\email{willettk@carleton.edu}

\and

\author{Bruce G. Elmegreen}
\affil{IBM T. J. Watson Research Center, PO Box 218, Yorktown Heights,
New York 10598 USA}
\email{bge@watson.ibm.com}

\and

\author{Deidre A. Hunter}
\affil{Lowell Observatory, 1400 West Mars Hill Road, Flagstaff, Arizona
86001 USA}
\email{dah@lowell.edu}

\begin{abstract}

Fourier transform power spectra of major axis cuts in V and \ha\
images were made for a sample of 9 irregular galaxies. These power
spectra reveal structure over a wide range of scales. For 6 of the
galaxies the power spectrum slopes at intermediate scales (10--400
pc) in the V-band images range from $-1.3$ to $-1.5$. 
The similarity of slopes suggests that the same processes are structuring 
these systems.
These slopes are 
slightly shallower than what is observed in other galaxies in \HI,
molecular emission, dust extinction, and optical light. Three of
the galaxies have flat power spectra like noise from the sky;
these three galaxies are relatively indistinct in the direct
images. The power spectrum slope for \ha\ steepens with increasing
star formation rate, ranging from a shallow value comparable to
the noise at low rates to a steep value with a slope of $\sim-1.5$
at high rates. This change reflects the increasing areal filling
factor of \ha\ emission with increasing star formation rate, and
an apparently universal slope inside the \ha\ regions that is
comparable to that for Kolmogorov turbulence. The power spectrum
of HI in one galaxy has a steeper power law, with a slope of
$\sim-2.9$.
The fact that the
power laws of star formation are about the same for dwarf galaxies
and giant spiral galaxies suggests the microscopic processes are
the same, independent of spiral density waves and galaxy size.
\end{abstract}


\section{Introduction} \label{sec-intro}

Fourier transform power spectra of gas and stars in a variety of
galaxies have shown power laws over a factor of $\sim100$ in
scale. This includes observations of local HI (Crovisier \& Dickey
1983; Green 1993; Dickey et al. 2001) and  CO emission (Stutzki
\et\ 1998), local HI absorption (Deshpande, Dwarakanath, \& Goss
2000), HI emission from the Small (Stanimirovic et al. 1999) and
Large Magellanic Clouds (Elmegreen, Kim, \& Staveley-Smith 2001),
dust extinction in the Small Magellanic Cloud (Stanimirovic et al.
2000), galactic nuclei (Elmegreen, Elmegreen \& Eberwein 2002),
and optical light from nearby galaxies (Elmegreen, Elmegreen \&
Leitner 2003a; Elmegreen \et\ 2003b). Other evidence for fractal
structure in the interstellar gas comes from perimeter-area
relations (e.g., Falgarone, Phillips \& Walker 1991) and
box-counting methods (e.g., Westpfahl \et\ 1999).

The most likely cause of this scale-free structure is a
combination of gravitational collapse, compressible MHD
turbulence, and nested shells from multiple supernovae (see review
in Elmegreen \& Scalo 2004). Cluster formation and star formation
occurs in the compressed regions (see reviews in Mac Low \&
Klessen 2004; Klessen 2004).

Although the technique of using Fourier transform power spectra to
analyze structure on galactic scales has been widely used, there
has been no previous work on irregular (Im) galaxies outside the
Local Group. Availability of images in multiple passbands of a
large sample of irregular galaxies (Hunter, Elmegreen \& van
Woerden 2001; Hunter \& Elmegreen 2004, in preparation) now gives
us the opportunity to carry this type of analysis to a larger
number of Im systems. Thus we undertook a study of the Fourier
transform power spectra of a sample of Im and Blue Compact Dwarf
(BCD) galaxies. The irregulars previously examined
---the LMC and SMC (Stanimirovic \et\ 1999; Elmegreen \et\ 2001)---offer
higher spatial resolution than our images, but our galaxies offer
a more diverse sampling of the Im class. Comparison of this sample
to the LMC, SMC, and spiral galaxies allows us to determine
whether Im galaxies as a group are homogeneous and to examine
similarities and differences with larger galaxies. Spectra that
follow a power law have resembled the power spectrum of velocity
for Kolmogorov turbulence in an incompressible medium.  The
reasons for this similarity are unknown (see discussion in
Elmegreen, Elmegreen \& Leitner 2003a).  Deviations from the
Kolmogorov spectrum in different galaxies may indicate different
types of turbulent activity or excitation mechanisms.

This paper presents Fourier transform power spectra of
one-dimensional linear cuts along the major axes of nine irregular
galaxies. The data include V-band and \ha\ images and one \HI\
map. Optical passbands give information about stellar structure
with a large range of spatial scales; \ha\ tracks the current star
formation; and \HI\ shows the distribution of diffuse gas.

\section{Galaxy Data} \label{sec-galaxies}

The V-band and \ha\ images of our sample galaxies were obtained as
part of a larger survey of the star-forming properties of 140 Im
galaxies. The \ha\ observations are described by Hunter \&
Elmegreen (2004). These images were obtained through narrow-band
(FWHM$\sim$30 \AA) filters and a continuum-only off-band filter
centered at 6440 \AA\ with a FWHM of 95 \AA. Most were obtained
using the Perkins 1.8 m telescope at Lowell Observatory and an
800$\times$800 TI CCD. Large galaxies were imaged in multiple
fields that were mosaicked into a single image. Usually multiple
images through the \ha\ filter and sometimes also through the
off-band filter were obtained and combined to remove cosmic rays.
The off-band image was shifted, scaled, and subtracted from the
\ha\ image to remove the stellar continuum.
The pixel scale (given in Table \ref{tab-properties})
for the \ha\ image of DDO 50 was 0.433\arcsec\
and the rest were 0.488\arcsec.
The final field of view of the \ha\ images was 5.5\arcmin\ for
DDO 43, DDO 88, NGC 1156, NGC 1569, NGC 3738, and VIIZw 403.
The field of view of the DDO 50 image was 9.7\arcmin; for
DDO 133, 6.3\arcmin; and for NGC 2366, 12.5\arcmin.
The resulting seeing (FWHM of an isolated star) on these images
was 1.6\arcsec--1.8\arcsec\ for DDO 43, DDO 50, NGC 1156,
and VIIZw 403; the rest had a seeing of 2.2\arcsec--2.6\arcsec.

V-band observations are described by Hunter \& Elmegreen (in
preparation). The images of NGC 2366 are described in detail by
Hunter, Elmegreen, \& van Woerden (2001). Most V images were
obtained with the Lowell Observatory 1.1 m Hall telescope using a
SITe 2048$\times$2048 CCD.  VIIZw 403 was observed at the Perkins
1.8 m telescope, and images of NGC 1569 and NGC 2366 were obtained
for us by P. Massey using the Kitt Peak National Observatory (KPNO) 4 m
telescope and a Tektronix 2048$\times$2048 CCD. Several images
were obtained of each galaxy with offsets of
20\arcsec---30\arcsec\ to improve the final flat-fielding. The
electronic pedestal was subtracted using the overscan strip, and
the images were flat-fielded using sky flats. Foreground stars and
background galaxies were edited out of the images and the
background sky was fitted with a two-dimensional Legendre function
and subtracted.
Pixel scales are given in Table \ref{tab-properties}.
The NGC 1569 and NGC 2366 V-band images, taken at KPNO,
had a pixel scale of 0.42\arcsec, a field of view of 14.3\arcmin,
and a seeing of 1.7\arcsec\ and 2.2\arcsec.
The image of VIIZw 403, taken with the Perkins 1.8 m telescope,
had a pixel scale of 0.608\arcsec, a field of view of 4.8\arcmin,
and a seeing of 2.1\arcsec.
The rest of the galaxies, imaged with the Hall 1.1 m telescope,
had a pixel scale of 1.134\arcsec, a field of view of 17\arcmin,
and a seeing of 2.8\arcsec--3.4\arcsec.

An integrated \HI\ map of NGC 2366 was also analyzed. These data are a
combination of VLA C and D array interferometric observations
(Hunter \et\ 2001), and the map is a flux-weighted moment zero
image made from the data cube.  The beam size of the \HI\ map is
33.6\arcsec$\times$28.9\arcsec, and this represents the resolution
of the map. The pixel size is 5\arcsec, and the map is
42.7\arcmin\ on a side.
This map is sensitive to structures up to 15\arcmin\ in size.

The center of the galaxy and position angle of the major axis
are taken from Hunter \& Elmegreen (in preparation).
They determined these parameters and the ellipticity of the galaxy from
an outer contour of the V image block-averaged by factors of a few
to increase signal-to-noise in the fainter outer parts.
The center is simply the geometrical center of
this outer isophote and the major axis is the longest bisector
that passes through the center
that, as much as possible, symmetrically divides the galaxy.
A single center and major axis were used for all passbands
of a galaxy.

Foreground stars and background galaxies were edited
from the V-band images before surface photometry was performed
in fixed ellipses that increase in semi-major axis length
in approximately 10\arcsec\ steps.
The surface photometry of the V-band image was fit
with $\mu = \mu_0 + 1.086 r/{\rm R}_D$ which represents an exponential disk
and R$_D$ is the scale-length of the disk.
All of the galaxies discussed here show an exponential fall-off
in V-band surface photometry, and the R$_D$ are given in
Table \ref{tab-properties}.

\section{Data Analysis} \label{sec-analysis}

Table \ref{tab-properties} shows a summary of the global
properties of the galaxies in our sample. Galaxies were selected
based on their diverse sizes and star formation rates as well as
their large apparent sizes relative to the pixel scale of the CCD.
Our choice of galaxies was also restricted to those that were not
heavily resolved into point sources, which would have distorted
the power spectra at high spatial frequencies. Foreground stars
and background galaxies in the images were removed using a routine
that interpolates across a circle centered on the object being
removed.

The images were processed and transformed into power spectra using
a combination of the Image Reduction and Analysis Facility (IRAF)
software package and a FORTRAN program of our construction. The
FITS images were transformed into ASCII arrays with values
corresponding to the intensities of each pixel. The FORTRAN
program then extracted linear scans of the array along the major
axes of the galaxy.
Ten strips
were chosen parallel to the major axis of each galaxy and
separated by one pixel. Regularly-spaced points along each strip
were taken to have intensity values equal to the values of the
pixels in which the points lay, without interpolation. The
regions over which the strips were taken are outlined on the V and
\ha\ images in Figure \ref{fig-strips} and on the \HI\ map of NGC
2366 in Figure \ref{fig-histrip}.

Exponential disks were removed from the V and
\ha\ images by dividing each intensity strip by
the average exponential profile:
\begin{equation}
I_{cor} (n) = I(n) / (10^{-0.4 * b * |n_c - n|}),
\label{eq:expdisk}
\end{equation}
where $I(n)$ is the observed count in the $n$th pixel out to the sky
limit, $n_c$ is the pixel number of the center of the galaxy, and
$b$ describes the exponential fall-off of the disk (determined at
V-band). 
For the \HI\ map we normalized with an exponential fit to
the \HI\ surface density radial profile.
In Figure \ref{fig-exp} we show the result to one power spectrum
of removing the underyling exponential disk. One can see that 
the exponential disk contributes to a steepening at higher relative
spatial frequency.

The choice of linear scans rather than azimuthal was influenced by
the lack of obvious symmetry in many of the galaxies, as well as
concerns that deprojecting the galaxy would introduce
interpolation errors at high spatial frequencies. The linear scans
suffer from edge effects, however, as dissimilar values at the two
ends cause a broad erroneous response in the Fourier transform. To
remove these edge effects, a cosine taper with a length of 20
pixels was added to each strip end. When fitting the entire
spectrum with a power law, tapering of the ends changed the
overall slope by less than 0.1 for each image tested, mitigating an
observed turn-up at the lowest $k$. These lowest $k$ spatial frequencies
were not considered in our final analysis since they reflect
effects from galaxy disk gradients.

Several of the galaxies have sparse H$\alpha$ and the linear cuts
do not go through the brightest emission regions even though they
go through the main part of the V-band emission. These cases also
have noisy power spectra for H$\alpha$ that do not show clear
power laws. This situation leads to our eventual conclusion that
the power spectrum of H$\alpha$ emission resembles a Kolmogorov
power law wherever the filling factor of H$\alpha$ emission is
high enough to see the turbulent structure above the noise; this
tends to occur in the galaxies with the highest star formation
rates per unit area. This conclusion is true even though other
cuts, specially chosen for the faint cases, might show power laws
over small regions. We considered taking 2D power spectra for our
survey to include all of the emission, but then the edge effects
become more prominent (a high fraction of the area is at the edge)
and the sparse galaxies are still dominated by noise in 2D.  Our
choice in the end was to be systematic rather than selective,
taking major axis cuts over a broad swath (10 pixels wide) to
include as much of the emission as possible.

Fourier sine and cosine transforms were applied to the intensity
strips,
\begin{equation}
I_s(k) = \sum_{n=1}^N \sin(2kn\pi /N)\times I_{cor} (n) ,
\label{eq:cosft}
\end{equation}

\begin{equation}
I_c(k) = \sum_{n=1}^N \cos(2kn\pi /N)\times I_{cor} (n) .
\label{eq:sinft}
\end{equation}
In these equations, $N$ is the total number of pixels in the
intensity scan, given in Table \ref{tab-n}, and $k$ is the wavenumber.
The power spectrum is
computed by summing the squares of the individual sine and cosine
transforms:
\begin{equation}
P(k) = I_s(k)^2 + I_c(k)^2 . \label{eq:power}
\end{equation}
The final power spectra used here are the averages of the power
spectra obtained for each of the 10 scans along the major axis.

The power spectra were plotted in log-log space as a function of
their relative spatial frequency RSF=$k/\left(N/2\right)$. The RSF
is the spatial frequency normalized to the highest value that can
be measured, $N/2$. Values of RSF=1 correspond to the smallest
full sine wave of structure, having a wavelength of 2 pixels.
(Note that our convention is to include $2\pi$ explicitly inside
the argument of the sine and cosine functions, so wavelength is
$1/k$).

Linear fits to the average power spectra are taken for RSF between
0.03 and 0.8. High $k$ is avoided because stars and other point
sources contaminate the power spectrum there. Seeing was
generally 2\arcsec--3\arcsec\ FWHM (2--3 pixels) for the V-band images,
corresponding to $k\sim1/2.5=0.4$. Low $k$ is avoided because
they are affected by galaxy gradients
and the imposed cosine taper.  The
cosine taper has a wavelength of 40 pixels, and so contributes a
small spike in the power spectrum at $k=1/40=0.025$.

\section{Results} \label{sec-results}

\subsection{V-band Images}

Power spectra for the 9 galaxies observed in V-band are shown in
Figure \ref{fig-vband}. The lengths of the power spectra vary with
the number of pixels in the strip. The spectra are offset from
each other for clarity. Sky noise is shown at the bottom. Sky
noise comes from 10 strips outside of NGC 2366, which give an
average slope of $-0.3$. All of the images have about the same sky
power spectrum, with slopes varying by $\pm 0.2$.  The reference
line in the lower left corner has a slope of $-5/3$, which would
be the slope of the power spectrum for idealized Kolmogorov
turbulence.

The power spectra along the V-band major axes vary widely for our
sample. We estimated the intrinsic variations by determining the
power spectra of many strips with different position angles
through the center of NGC 1156. This is a round galaxy in the
outer parts with no
outstanding characteristics in any direction.  The standard
deviation of power spectrum slopes was 0.2.  Differences in strip
alignment and variations within a galaxy suggest the uncertainty
in any individual slope is $\pm 0.3$. We take this to be our
uncertainty for all of the power spectra presented here.

Table \ref{tab-slopes} gives the slopes of lines fit to the power
spectra.
The spectra of DDO 43, DDO 50, NGC 1569, and NGC 3738 were determined
by eye to roughly follow a two-part power law; the
``Fit Division'' column gives the dividing point at which the two
slopes were measured.
``Low RSF'' are fits to the portion of the power spectrum less than
the ``Fit Division'' RSF, while ``High RSF'' are fits to the portion
greater than the division.
NGC 2366 is fit most closely by a three-part power law; both
divisions are given in the second column.
Overall, the galaxies' slopes divide into two groups:
$\sim-1.3$ and $\sim-0.5$. All slopes in the V-band are flatter
than the Kolmogorov value of $-1.66$.

NGC 2366 and NGC 3738 have clear power laws with relatively few
features. DDO 43, DDO 50, DDO 133, and NGC 1569 also have power
laws but with more random deviations. The bump in NGC 1569's
spectrum at RSF$\sim0.1$ is from the two super-star clusters near
the center of the galaxy (O'Connell, Gallagher, \& Hunter 1994;
Hunter \et\ 2000); when the clusters are removed, the bump disappears.

DDO 88, NGC 1156, and VIIZw 403 have flat power spectra, resembling
noise. The 10 individual spectra that went into these averages
have larger deviations than they do for the other galaxies.
VIIZw 403 has a two-part exponential disk with a steeper
exponential in the center --- not uncommon for BCDs.
However, NGC 3738 also has a two-part exponential disk and
does not show the low frequency features of VIIZw 403. DDO 88 has a
relatively high sky-to-galaxy ratio in the image used. All three
galaxies with flat power spectra are round and relatively
homogenous in V.  Presumably these power spectra are measuring
cluster and pixel noise in the images, rather than any large scale
star formation patches.

The power spectrum slope of the V-band images does not correlate
with galaxy absolute magnitude M$_V$, central surface brightness
$\mu_V^0$, disk scale length R$_D$, or average surface brightness
within R$_D$. To examine the importance of shear, we determined
the radii at which the solid body parts of the rotation curves
ended and normalized these radii to the disk scale lengths (Hunter
\et\ 2001; Simpson \et\ 2004a,b; Stil \& Israel 2002; Swaters
1999; McIntyre 2003). Rotation curves of DDO 133, NGC 3738, or
VIIZw 403 were not available. We found no correlation between the
power spectra and these relative radii for shear.

\subsection{H$\alpha$ Images}

Power spectra of \ha\ images also have a variety of shapes, as
shown in Figure \ref{fig-haband}. The slopes are given in Table
\ref{tab-slopes}. The average lengths of the \ha\ strips were
larger by a factor of $\sim$1.5 than the lengths for the V-band
because the \ha\ pixels were smaller. Power spectra of sky noise
in the \ha\ images varied from $0$ to $-0.2$, with an average of
$-0.1\pm0.06$.  The sky noise shown in Figure \ref{fig-haband} has
this slope and is from NGC 2366.

The relative scale for the power spectra of V-band and \ha\ are
the same;  absolute power is not shown in Figures \ref{fig-vband}
and \ref{fig-haband}. The scale of distance relative to the RSF
depends on the length of the major axis cut measured in pixels;
these are given in Table \ref{tab-n}. Absolute sizes of individual
features can be calculated using the original dimensions of the
strip. Power-law behavior is independent of the pixel size of the
image.

The large variety for the \ha\ power spectra is related to the
great diversity of the \ha\ images. Several galaxies, such as NGC
1569 and NGC 2366, have many large \HII\ regions and steep power
spectra, while others, such as DDO 88 and DDO 50, have sparse
\HII\ regions along the extracted strip and flat power spectra
resembling noise. When the HII regions are sparse, other intensity
strips that pass through individual emission regions could have
been analyzed instead of the major axis strips used here. These
emission regions are probably turbulent like most interstellar
gas, and their power spectra might have shown the appropriate
power spectra if they had enough spatial resolution, but here they
contain far too few pixels to give sensible power spectra by
themselves, and in 2D maps they are only a small fraction of the
total structure, which is still dominated by noise in these cases.

NGC 1569 follows a two-part power law with a ledge at
RSF$\sim0.1$, corresponding to a scale of $\sim$190 pc. At low
RSF, the slope for NGC 1569 is $-1.7$ and at high RSF the slope is
$-0.9$. NGC 3738 also has a two-part power spectrum, with a slope
at low RSF of $-1.0$ and flat slope at high RSF. NGC 1156 has many
large \HII\ regions and a relatively straight power spectrum with
a slope of $-1.1$. It flattens at high RSF, presumably because of
unresolved HII regions and pixel noise. The NGC 2366 power
spectrum is dominated by the supergiant \HII\ region NGC 2363 in
the SE corner; few other details in the power spectrum are
visible.

DDO 43, DDO 50, DDO 88, and VIIZw 403 all have sparse \ha\
distributions along the extracted strip
and flat power spectra. The sources of ringing for
DDO 43 and VIIZw 403 are unclear, as there are few \ha\ features.
In general, small \HII\ regions appear as point sources and
contribute to the flattening of the spectra at high frequency.

Figure \ref{fig-ha_sfr} shows that the power spectrum slope for
\ha\ steepens with increasing star formation rate (as obtained
from Hunter \& Elmegreen 2004). The exceptional case, NGC 2366,
would apparently have a steeper power spectrum without
contamination from NGC 2363.

\subsection{\HI\ Map of NGC 2366}

Figure \ref{fig-hiband} shows the power spectrum of \HI\ from NGC
2366. \HI\ images were also available for DDO 43 and DDO 88, but
the number of pixels was too small to give a useful result. The
spectrum of NGC 2366 follows a power law reasonably well for
RSF$<0.5$, with a steep slope of $-2.9$ for the 1D strip. The bump
at RSF$=0.1$ may result from a neutral cloud associated with the
supergiant \HII\ complex NGC 2363, or it may result from an HI
ring seen in projection (Hunter \et\ 2001). The HI strip passes
through both the edge of NGC 2363 and one part of the ring. Power
spectra of \HI\ strips in the LMC have a slope that is shallower
than that for NGC 2366 by about 1 (Elmegreen, Kim, \&
Staveley-Smith 2001).  The reason for this difference is not
known. The LMC observation contains single beam data in addition
to interferometric data, while NGC 2366 only has interferometric
data.

\section{Summary} \label{sec-summary}

Fourier transform power spectra of major axis cuts in V and
\ha-band images of a sample of 9 irregular galaxies were examined.
On scales of $10-400$ pc, 6 galaxies have V-band power spectra
with power law slopes of $-1.4\pm0.3$. This similarity suggests
that the same processes are structuring star formation regions in
these systems. Three galaxies, which are all relatively round and
indistinct, have flat power spectra similar to noise.

The \ha\ images have power spectra with slopes that steepen for
greater star formation rates.  This is presumably the result of a
greater angular filling factor for \ha\ at higher star formation
rates, along with a near universal intrinsic power spectrum for
\ha\ and other interstellar regions that is close to that for
Kolmogorov turbulence.  At low \ha\ filling factors, the power
spectrum is dominated by noise.

An \HI\ map of NGC 2366 yielded a power spectrum with a relatively
steep slope, but this could be the result of low-frequency
contamination by a giant HI ring and part of the giant HI cloud
associated with NGC 2363.

Differences between the V-band and H$\alpha$ power spectra are
usually the result of the influence of a few bright regions that
show up in one band and not the other. These differences are to be
expected for small galaxies where the number of emission regions
is small and the age of each one differs.  In larger galaxies
there is a wider sample of structure at each age so particular
regions do not stand out in the power spectra. Even so, there
could be a difference between V-band and H$\alpha$ power spectra
in large galaxies.  If, for example, the largest regions form
stars for the longest times and have the weakest H$\alpha$
emission compared to optical (i.e., they have lower H$\alpha$
equivalent widths), then the power spectrum for V-band would be
steeper than for H$\alpha$. Alternatively, the stars that make up
the V-band could diffuse to a more uniform structure over time
while the turbulence seen in H$\alpha$ is always rejuvenated; then
the power spectrum for V-band would be shallower than for
H$\alpha$. These possible differences are not expected to show up
in small galaxies where a few individual sources dominate.

Power spectra give new insight into galactic structure. For images
with a large number of pixels and few foreground stars, the power
spectra of the azimuthal profiles and major-axis cuts are all
approximately power laws when the emission is dominated by star
formation. The power spectra resemble noise when the emission is
dominated by old dispersed clusters and field stars.  This result
implies that star formation in the optical and \ha\ passbands is
approximately scale-free, which means there is no characteristic
mass or luminosity for OB associations and star complexes.  The
scale-free nature also implies that the gas processes leading to
star formation are scale-free, as would be the case for
gravitational instabilities (Semelin et al. 1999) and turbulence
compression (Elmegreen \& Scalo 2004).  Most likely a combination
of self-gravity and turbulence structures the gas, triggering star
formation. The driving force for the turbulence may be
self-gravity as well, with a substantial contribution from
supernovae and other young-stellar pressures. The fact that the
power laws of star formation are about the same for dwarf galaxies
and giant spiral galaxies suggests the microscopic processes are
the same, independent of spiral density waves and galaxy size.

\acknowledgments

KWW would like to thank Kathy Eastwood and the 2004
Research Experience for Undergraduates program at Northern Arizona
University which is funded by the National Science Foundation
under grant 9988007.
Funding for this work was also provided
by the National Science Foundation through grants AST-0204922 to DAH
and AST-0205097 to BGE.

\clearpage

\begin{deluxetable}{lccccccc}
\tabletypesize{\scriptsize}
\tablecaption{Galactic properties. \label{tab-properties}}
\tablewidth{0pt}
\tablehead{
\colhead{} & \colhead{} & \colhead{} &
\multicolumn{2}{c}{Pixel Size} & \colhead{} & \colhead{} & \colhead{}
\\
\cline{4-5}
\colhead{} & \colhead{} & \colhead{Distance\tablenotemark{a}}
& \colhead{V-band\tablenotemark{b}}
& \colhead{\ha\ \tablenotemark{c}} & \colhead{R$_D$\tablenotemark{b}}
& \colhead{log SFR$_D$\tablenotemark{c}}
& \colhead{Avg. surface brightness\tablenotemark{b}}
\\
\colhead{Galaxy} & \colhead{Type\tablenotemark{a}}
& \colhead{[Mpc]} & \colhead{[arcsec]}
& \colhead{[arcsec]}
& \colhead{[kpc]} & \colhead{[\sfrunit]} & \colhead{[mag arsec$^{-2}$]}
}
\startdata
DDO 43 & Im & 5.5 & 1.134 & 0.488 & 0.43 & -2.19 & 22.01 \\
DDO 50 & Im & 3.4 & 1.134 & 0.433 & 1.11 & -1.83 & 21.65 \\
DDO 88 & Im & 7.4 & 1.134 & 0.488 & 0.77 & -2.60 & 21.60 \\
DDO 133 & Im & 6.1 & 1.134 & 0.488 & 2.15 & -2.93 & 23.14 \\
NGC 1156 & IB(s)m & 7.8 & 1.134 & 0.488 & 0.82 & -0.87 & 19.02 \\
NGC 1569 & IBm & 2.5 & 0.420 & 0.488 & 0.28 & \x0.11 & 17.14 \\
NGC 2366 & IB(s)m & 3.2 & 0.420 & 0.488 & 1.28 & -1.73 & 21.77 \\
NGC 3738 & Im & 4.9 & 1.134 & 0.488 & 0.77 & -1.72 & 9.63 \\
VIIZw 403 & BCD(Pec) & 4.4 & 0.608 & 0.488 & 0.52 & -1.82 & 21.82 \\
\enddata
\tablenotetext{a}{See Hunter \& Elmegreen 2004 for references.}
\tablenotetext{b}{From Hunter \& Elmegreen, in preparation.
The average surface brightness is that in V-band within R$_D$.}
\tablenotetext{c}{From Hunter \& Elmegreen 2004. SFR$_D$ is star formation rate
normalized to the size of the galaxy as given by $\pi {\rm R}_D^2$.}
\end{deluxetable}

\clearpage

\begin{deluxetable}{lcccccc}
\tablecaption{Strip lengths\tablenotemark{a}
\label{tab-n}}
\tablewidth{0pt}
\tablehead{ \colhead{} &
\colhead{$N_V$} &
\colhead{$L_V$} &
\colhead{$N_{H\alpha}$} &
\colhead{$L_{H\alpha}$} &
\colhead{$N_{HI}$}
& \colhead{$L_{HI}$} \\
\colhead{Galaxy}
& \colhead{(pixel)}
& \colhead{(arcmin)}
& \colhead{(pixel)}
& \colhead{(arcmin)}
& \colhead{(pixel)}
& \colhead{(arcmin)}
}
\startdata
DDO 43 & 192 & 3.6 & 617 & 5.0 & \nodata & \nodata \\
DDO 50 & 431 & 8.1 & 937 & 6.8 & \nodata & \nodata \\
DDO 88 & 370 & 7.0 & 647 & 5.3 & \nodata & \nodata \\
DDO 133 & 325 & 6.1 & 550 & 4.5 & \nodata & \nodata \\
NGC 1156 & 279 & 5.3 & 555 & 4.5 & \nodata & \nodata \\
NGC 1569 & 702 & 4.9 & 660 & 5.4 & \nodata & \nodata \\
NGC 2366 & 1267 & 8.9 & 809 & 6.6 & 106 & 8.8 \\
NGC 3738 & 254 & 4.8 & 225 & 1.8 & \nodata & \nodata \\
VIIZw 403 & 471 & 4.8 & 657 & 5.3 & \nodata & \nodata \\
\enddata
\tablenotetext{a}{$N$ is the number of pixels in the strip
extracted from the V-band, H$\alpha$, or HI image; $L$ is the
length of that strip in arcminutes.}
\end{deluxetable}

\clearpage

\begin{deluxetable}{lccccccccccc}
\rotate
\tabletypesize{\scriptsize}
\tablecaption{Slopes of power
spectra. \label{tab-slopes}}
\tablewidth{0pt}
\tablehead{
\colhead{} & \multicolumn{5}{c}{V} & \colhead{} &
\multicolumn{4}{c}{\ha\ } & \colhead{\HI\ }
\\
\cline{2-6} \cline{8-11}
\colhead{Galaxy}
& \colhead{Fit division\tablenotemark{a}} &
\colhead{Low RSF\tablenotemark{a}} & \colhead{Mid RSF} & \colhead{High RSF} &
\colhead{Total\tablenotemark{b}}  &
\colhead{} &
\colhead{Fit division\tablenotemark{a}} &\colhead{Low RSF} & \colhead{High RSF} &
\colhead{Total\tablenotemark{b}} &
\colhead{Total\tablenotemark{b}}
}
\startdata
DDO 43 & 0.2 & -2.0 & \nodata & -0.7 & -1.4 & & \nodata &
\nodata & \nodata & -0.5 & \nodata \\
DDO 50 & 0.4 & -1.1 & \nodata & -1.9 & -1.3 & & \nodata &
\nodata & \nodata & -0.1 & \nodata \\
DDO 88 & \nodata & \nodata & \nodata & \nodata & -0.3 & & \nodata & \nodata & \nodata & -0.0 & \nodata \\
DDO 133 & \nodata & \nodata & \nodata & \nodata & -1.5 & & \nodata & \nodata &
\nodata & -0.3 & \nodata \\
NGC 1156 & \nodata & \nodata & \nodata & \nodata & -0.3 &
& \nodata & \nodata & \nodata & -1.1 & \nodata \\
NGC 1569 & 0.2 & -1.7 & \nodata & -0.2 & -1.3 & & 0.1 & -1.7 & -0.9 &
-1.5 & \nodata \\
NGC 2366 & 0.05, 0.3 & -1.0 & -2.0 & -0.7 & -1.4 &  & 0.3 &
-1.5 & -0.3 & -1.6 & -2.9 \\
NGC 3738 & 0.08 & -2.9 & \nodata & -0.8 & -1.3 & & 0.1 & -1.0 & -0.2 &
-0.7 & \nodata \\
VIIZw 403 & \nodata & \nodata & \nodata & \nodata & -0.4 & &
\nodata & \nodata & \nodata & -0.2 & \nodata \\
\enddata
\tablenotetext{a}{Divisions in Relative Spatial Frequency (RSF) for galaxies
whose power spectra were fit with multi-part power laws.}
\tablenotetext{b}{``Total'' is the slope fit to the power spectra between RSF of
0.03 and 0.8}
\end{deluxetable}

\clearpage

\begin{figure}
\caption{ V-band (left) and \ha\ images (right) of the irregular
galaxies examined in this study. The rectangle in each image
outlines the area where single pixel-wide linear strips were taken
and used to produce an average one-dimensional power spectrum. The
scale of the image is shown in the bottom left-hand corner. In
most images North is at the top and East to the left. However,
some images were rotated 90\arcdeg\ to more easily facilitate
taking the strips and in those images North is to the left and
East is down. Most foreground stars and background galaxies have
been edited from the images. Images are from Hunter \& Elmegreen
(2004 and in preparation). The galaxies are, from top to bottom:
(a) DDO 43, DDO 50, DDO 88; (b) DDO 133, NGC 1156, NGC 1569; and (c)
NGC 2366, NGC 3738, and VIIZw 403. \label{fig-strips}}
\end{figure}

\epsscale{1.2}

\epsscale{1.2}

\epsscale{1.2}

\clearpage

\begin{figure}
\epsscale{1.0}
\caption{
Integrated \HI\ map of NGC 2366 from Hunter \et\ (2001). The white rectangle
outlines the area where pixel-wide
linear strips were taken
and used to produce an average
one-dimensional power spectrum.
The beam of the map is 33.6\arcsec\ $\times$ 28.9\arcsec\ and
the pixel size is 5\arcsec.
\label{fig-histrip}}
\end{figure}

\clearpage

\begin{figure}
\epsscale{0.8} 
\caption{ Power
spectra in the H$\alpha$-band for NGC 1156. In ``Without exponential
disk'' the underlying exponential disk has been removed from the
extracted strips before the power spectrum is formed.
In ``With exponential disk'' the underlying exponential disk
has been left untouched.
The two spectra are offset from each other in the y-axis to allow comparison. 
The scale at the bottom is the normalized
wavenumber $k/\left(N/2\right)$ for strip length $N$. $N$ is given 
in Table \protect\ref{tab-n}. 
\label{fig-exp}}
\end{figure}

\clearpage

\begin{figure}
\epsscale{0.75} 
\caption{ Power
spectra in the V-band for the nine galaxies surveyed. Spectra are
offset from each other in the y-axis to allow comparison of
relative features. The scale at the bottom is the normalized
wavenumber $k/\left(N/2\right)$ for strip length $N$.  
$N$ are given 
in Table \protect\ref{tab-n}. 
The
spectrum at the bottom labeled ``Sky Noise'' is the average sky in
the NGC 2366 image, and represents the characteristic flatness of
the noise power spectra in all of our galaxies. The dashed line
given for scaling shows the slope of $-5/3$ seen in ideal
Kolmogorov turbulence. \label{fig-vband}}
\end{figure}

\clearpage

\begin{figure}
\epsscale{0.8} 
\caption{ Power
spectra in \ha\ for the nine galaxies surveyed, plotted as in
Figure 4. \label{fig-haband}}
\end{figure}

\epsscale{0.9}

\begin{figure}
\epsscale{1.0} 
\caption{ Star
formation rate per unit area (SFR$_D$) vs. the slope of the power
spectrum in \ha. SFR$_D$ is calculated from the \ha\ luminosity
according to the formula given in Hunter \& Elmegreen (2004)
and normalized to the area $\pi$R$_D^2$.
The slope for NGC 2366 is skewed due to the presence of the giant
\HII\ region NGC 2363 and does not give a complete representation
of structure at all scales in the galaxy. \label{fig-ha_sfr}}
\end{figure}

\begin{figure}
\epsscale{0.8}
\caption{
Power spectrum in \HI\ for NGC 2366.
The scale at the bottom is the normalized wavenumber at $k$ = 1/(2 pixel).
A relative spatial frequency of 1 represents the center of the power spectrum.
The dashed line given for scaling shows the slope of $-5/3$ seen in
ideal Kolmogorov turbulence.
\label{fig-hiband}}
\end{figure}

\end{document}